\documentclass[%
 reprint,
 amsmath,amssymb,
 aps,
]{revtex4-2}

\usepackage{graphicx}% Include figure files
\usepackage{dcolumn}% Align table columns on decimal point
\usepackage{bm}% bold math
\DeclareMathOperator{\tr}{\text{tr}}

\begin{document}

\preprint{APS/123-QED}

\title{Quantum Data Compression and Quantum Cross Entropy}% Force line breaks with \\

\author{Zhou Shangnan}
\email{snzhou@stanford.edu}
\affiliation{Stanford Institute for Theoretical Physics, Stanford University, Stanford, CA 94305, USA}
\affiliation{GreenQubit AI, Cupertino, CA 95014, USA}

\date{November 10, 2023}% It is always \today, today,

\begin{abstract}
The emerging field of quantum machine learning has the potential of revolutionizing our perspectives of quantum computing and artificial intelligence. In the predominantly empirical realm of quantum machine learning, a theoretical void persists. This paper addresses the gap by highlighting the quantum cross entropy, a pivotal counterpart to the classical cross entropy. We establish quantum cross entropy's role in quantum data compression, a fundamental machine learning task, by demonstrating that it acts as the compression rate for sub-optimal quantum source coding. Our approach involves a novel, universal quantum data compression protocol based on the quantum generalization of variable-length coding and the principle of quantum strong typicality. This reveals that quantum cross entropy can effectively serve as a loss function in quantum machine learning algorithms. Furthermore, we illustrate that the minimum of quantum cross entropy aligns with the von Neumann entropy, reinforcing its role as the optimal compression rate and underscoring its significance in advancing our understanding of quantum machine learning's theoretical framework.
\end{abstract}

\maketitle

%\tableofcontents

\section{Introduction}

Machine learning, as a potent tool for pattern recognition in data, has garnered attention across diverse domains \cite{lecun2015deep, heaton2018ian}. With the evolution of quantum information science, there's a burgeoning curiosity in crafting machine learning algorithms tailored for quantum data and computers \cite{dunjko2016quantum, QMLNature, huang2021information}.

Though there have been numerous attempts to design quantum machine learning architectures \cite{lloyd2013quantum,KernelNature, KernelPRL, Barren, gao2018quantum, deepqnn, caro2022generalization, du2022quantum}, significant questions remain unanswered, primarily due to an underdeveloped theoretical base. A foundational pillar of classical machine learning is the classical cross entropy $H(p,q) = -\sum_i p_i \log q_i$. Its quantum equivalent, the quantum cross entropy $S(\rho, \sigma) = -\tr(\rho \log \sigma)$ \cite{max}, serves as an essential element in the theoretical bedrock of quantum machine learning.

At its core, machine learning revolves around pattern recognition and redundancy reduction. Similarly, data compression aims to encapsulate information in its most concise form, aligning it inherently with machine learning objectives. Hence, the problem of quantum data compression serves as a very representative and important problem of quantum machine learning. 

Data compression is also by itself an interesting and important topic. It enables more efficient data storage and faster data communication. The common classical data compression methods are run-length encoding \cite{robinson1967results}, variable-length coding \cite{salomon2007variable}, LZ77 \cite{ziv1977universal} and LZ78\cite{ziv1978compression} algorithms, and transform coding. Run-length encoding represents repetitive data using a single value and count; variable-length coding assigns shorter codes to more frequent symbols and longer codes to less frequent symbols; LZ77 and LZ78 exploits repeated sequences - the essence of these three methods is to find the repetitive or typical sequences. Transform coding involves converting a signal from one domain into another domain, and then often uses the above compression techniques to simplify the transformed data. 

In this letter, we draw connection between quantum data compression and quantum cross entropy $S(\rho, \sigma) = -\tr(\rho \log \sigma)$, offering an operational lens to interpret quantum cross entropy as the compression rate for sub-optimal quantum source coding. To demonstrate this, we integrate the principles of transform coding and variable-length coding with established findings in quantum data compression and quantum entropies \cite{qcoding, NielsenChuang, indeter, universal, stationary, hayashi2002quantum, ahlswede2003lossless, hayden2004structure, universal06}. Other related works include \cite{hayashi2010universal,rozema2014quantum, bellomo2017lossless}. Our protocol starts with constructing a computational basis based on our perceived information of the quantum source, denoted as $\sigma$. A unitary transformation is then applied to align the actual quantum source, $\rho$, with this computational basis. We subsequently project the source state onto a subspace defined by typical length and demonstrate the faithfulness of this projection. The state is ultimately reverted to its original basis post-projection. We show that the compression rate in this case is the quantum cross entropy $S(\rho, \sigma)$. This protocol's primary contribution lies in addressing more generalized sources. Furthermore, it captures the fundamental aspect of data compression, which is the exploitation of typicality.

Since data compression extracts the key features of the original source, it can be seen as a machine learning task. In this case, the quantum cross entropy serves as a loss function, whose minimum is the von Neumann entropy $S(\rho)$. This is consistent with the result that von Neumann entropy is the optimal compression rate. 

\section{Quantum Cross Entropy}

For discrete probability distributions $p$ and $q$ with the same support $\mathcal{X}$, the classical cross entropy is

\begin{equation}
    H(p,q) = -\sum_{x \in \mathcal{X}} p(x) \log q(x).
\end{equation}
We define the quantum cross entropy \cite{max} by extending the classical definition from probability distributions to density matrices. For two density matrices $\rho$ and $\sigma$, the quantum cross entropy is

\begin{equation}
    S(\rho, \sigma) = -\tr(\rho \log \sigma),
\end{equation}
if the support of $\rho$ is contained in the support of $\sigma$, and $+\infty$ otherwise.

Details of the properties of quantum cross entropy, and its relation to the maximum likelihood principle, is highlighted in \cite{max}.

\section{Information Source and Strong Typicality}
\label{strong}

There are different models in defining an information source. We start from one simple but fruitful model for a classical information source \cite{network}. The source emits a symbol from a finite alphabet $\mathcal{I}$ with $D$ symbols at each single use. We assume that different uses of the source are independent and identically distributed. A possible output from $N$ consecutive uses is a sequence $\textbf{x} = i_1, i_2, ... , i_n, ... , i_N$ sampled from $N$ random variables $I_1, I_2, ... , I_n, ... , I_N$. We denote the probability of emitting symbol $i$ on any given use of the source as $P(i) = p_i$. Typically, the frequency of occurrence of any given symbol $i$ in a sequence output is close to $p_i$. To formalize this intuition, we first define the empirical probability mass function of $\textbf{x}$ (also referred to as its type) as 

\begin{equation}
    \pi(i|\textbf{x}) = \frac{|\{n: i_n = i\}|}{N}, \ \text{for $i \in \mathcal{I}$}.
\end{equation}
For example, if $\textbf{x} = (0, 1, 1, 0, 0, 1, 0)$, then $\pi (i|\textbf{x}) = \frac{4}{7}$ for $i = 0$, and $\pi(i|\textbf{x}) = \frac{3}{7}$ for $i = 1$.

When $N$ is large, by the law of large numbers, for each $i \in \mathcal{I}$, 

\begin{equation}
    \pi(i|\textbf{x}) \xrightarrow{} P(i) \ \text{in probability}.
\end{equation}

We can then define the set of $\epsilon$-strong-typical $N$-sequences $\textbf{x}$ (or the strong typical set in short) as

\begin{equation}
    \mathcal{T}_{\epsilon}^{(N)} (I) = 
\{\mathbf{x}: |\pi(i|\textbf{x}) - P(i)| \leq \epsilon P(i) \ \text{for all } i \in \mathcal{I}
\}.
\end{equation}

Another useful concept is the set of $\epsilon$-weak-typical $N$-sequences $\mathbf{x}$ (or the weak typical set in short), which is defined as

\begin{equation}
    \mathcal{U}_{\epsilon}^{(N)} (I) = 
\Big\{\mathbf{x}: \Big|\frac{1}{N} \log \frac{1}{P(\textbf{x})} - H(p)\Big| \leq \epsilon
\Big\},
\end{equation}
where $H(p) = - \sum_{i \in \mathcal{I}} P(i) \log P(i)$ is the Shannon entropy, and $P(\textbf{x})$ is the probability that a certain sequence $\textbf{x}$ occurs.

A sequence that is $\epsilon$-strong-typical is definitely $\epsilon$-weak-typical, while the reverse does not always hold. Here is a list of properties \cite{network} that both strong and weak typical sequences possess. 

\textbf{1. Unit Probability Theorem.} Given $\epsilon > 0$. For any $\delta > 0$, when $N$ is sufficiently large, 

\begin{equation}
    P \big(\mathbf{x} \in
    \mathcal{T}_{\epsilon}^{(N)} (I) \big) \geq 1 - \delta.
\end{equation}

This means that as $N$ approaches infinity, the probability that a given sequence $\mathbf{x}$ is typical approaches one. 

\textbf{2. Probability of Each Sequence.} Given $\epsilon > 0$. For $N$ sufficiently large, if a sequence $\mathbf{x} \in \mathcal{T}_{\epsilon}^{(N)} (I) $, then
\begin{equation}
    2^{-N(H(\mathbf{x}) + \epsilon)} \leq P(\mathbf{x}) \leq 2^{-N(H(\mathbf{x}) - \epsilon)}.
\end{equation}

\textbf{3. Size of the Set.} Given $\epsilon > 0$ and $\delta > 0$, for $N$ sufficiently large, the number $| \mathcal{T}_{\epsilon}^{(N)} (I)|$ of $\epsilon$-typical sequences satisfies
\begin{equation}
    (1-\delta)2^{N(H(\mathbf{x})-\epsilon)} \leq | \mathcal{T}_{\epsilon}^{(N)} (I)| \leq 2^{N(H(\mathbf{x})+\epsilon)}.
\end{equation}

Now, we move on to the quantum case. The definition of a quantum information source \cite{NielsenChuang} we adopt is based on the idea that entanglement is what we are trying to compress and decompress. Formally, an identical, independently distributed (i.i.d.) quantum source is described by a Hilbert space $H$ and a density matrix $\rho$ on that Hilbert space, represented by $(\rho, H)$. We can view the state $\rho$ as part of a larger system which is in a pure state, and the mixed nature of $\rho$ is due to the entanglement between $H$ and the remainder of the system. At each use, a quantum source emits a quantum state that is on average $\rho$. After $N$ consecutive uses, the average output is $\rho^{\otimes N}$. We proceed to develop a quantum version of the strong typicality.

Suppose the density matrix $\rho$ can be decomposed as

\begin{equation}
    \rho = \sum_{i \in \mathcal{I}} P(i) |i\rangle \langle i|,
\end{equation}
where the $|i\rangle$'s form an orthonormal set, and $P(i)$'s are eigenvalues of $\rho$, which obey the same rules as a probability distribution. An $\epsilon$-strong-typical product state is a state  $|x\rangle = |i_1\rangle |i_2\rangle \cdots |i_N\rangle$ where $\mathbf{x} = i_1, i_2, ... , i_N$ forms a (classical) $\epsilon$-strong-typical sequence.

We define the $\epsilon$-strong-typical subspace $T(N, \rho, \epsilon)$ as the subspace spanned by all $\epsilon$-strong-typical product states. These product states form a basis of $T(N, \rho, \epsilon)$. The projector $Q(N, \rho, \epsilon)$ onto the subspace $T(N, \rho, \epsilon)$ is

\begin{equation}
\begin{aligned}
    Q(N, \rho, \epsilon) = \sum_{|x\rangle \ \epsilon \text{-strong-typical}} 
    &|i_1\rangle \langle i_1| \otimes \\
    &|i_2\rangle \langle i_2| \otimes \\
    &\cdots \otimes \\
    &|i_N\rangle \langle i_N|.
\end{aligned}
\end{equation}

We are able to extend the classical theorems of strong typicality to their quantum counterparts.

\textbf{1. Quantum Unit Probability Theorem.} Given $\epsilon > 0$. For any $\delta > 0$, when $N$ is sufficiently large,

\begin{equation}
    \tr \big( Q(N, \rho, \epsilon) \rho^{\otimes N} \big)
    \geq 1 - \delta.
\end{equation}

\textit{Proof.} 
\begin{equation}
\begin{split}
    \tr \big( Q(N, \rho, \epsilon) \rho^{\otimes N} \big)
&= \sum_{\mathbf{x} \  \epsilon \text{-strong-typical}} P(i_1) P(i_2) \cdots P(i_N) \\
&=  \sum_{\mathbf{x} \  \epsilon \text{-strong-typical}}  P(\mathbf{x}). \\
\end{split}
\end{equation}

When $\mathbf{x}$ is $\epsilon$-strong-typical, it is also $\epsilon$-weak-typical, and the result follows from the unit probability theorem of weak typicality.

\textbf{2. Dimension of the Subspace.} Given $\epsilon > 0$ and $\delta > 0$, for $N$ sufficiently large, the dimension $|T(N, \rho, \epsilon)|$  of $\epsilon$-strong-typical states satisfies
\begin{equation}
    (1-\delta)2^{N(S(\mathbf{\rho})-\epsilon)} \leq |T(N, \rho, \epsilon)|\leq 2^{N(S(\mathbf{\rho})+\epsilon)}.
\end{equation}
\textit{Proof.} When a state is $\epsilon$-strong-typical, it is also $\epsilon$-weak typical, and the theorem follows from the results in the classical case.

\section{Quantum Data Compression with Misinformed Sources}
\label{wrong}

We introduce a quantum data compression protocol that achieves asymptotically lossless compression, meaning that the fidelity of the compressed data approaches perfect losslessness as the number of quantum systems, $N$, becomes very large, even when our operational assumptions about the information source are not accurate.

Consider that we devise a compression-decompression scheme under the assumption that the quantum source is characterized by a density matrix $\sigma$ within a Hilbert space $H$. In contrast, the true nature of the quantum source is represented by $(\rho, H)$. Common discrepancies involve both the eigenvalues and eigenbases:

\begin{equation}
\begin{aligned}
    \sigma &= \sum_{i=1}^D q_i |a_i\rangle \langle a_i|, \quad
    \rho = \sum_{i=1}^D p_i |b_i\rangle \langle b_i|, \\
    \{q_i\} &\neq \{p_i\}, \quad
    \{|a_i\rangle \} \neq \{|b_i\rangle\}.
\end{aligned}
\end{equation}

This scenario deviates from the ideal situation where the source's attributes are accurately determined \cite{qcoding, NielsenChuang}. In such a case, a straightforward application of the projection onto the typical subspace of $\sigma$ proves to be ineffective, due to the absence of intersection between the typical subspaces $T(N, \rho, \epsilon)$ and $T(N, \sigma, \epsilon)$ for large values of $N$ and small $\epsilon$. Furthermore, the typical subspace $T(N, \sigma, \epsilon)$, which spans a dimension of $2^{N S(\sigma)}$, prescribes a compression rate corresponding to $S(\sigma)$. Should $S(\rho) > S(\sigma)$, it would indicate a compression rate below the optimal lossless threshold for the true quantum state $\rho$, hence leading to a degradation of fidelity.

\subsection{Reassessing the Classical Analogy}

To draw parallels and potentially glean insights, we consider the analogous classical case. Imagine a classical information source where the probability of emitting the $i$-th symbol is $p_i$, as opposed to the incorrectly assumed probability $q_i$. A simple approach to source coding would involve assigning to the $i$-th symbol a codeword with length $l_i = \log (1/q_i)$. The expected length $\langle l \rangle$ of a codeword in this scheme is given by

\begin{equation}
\label{eq:expected_length}
    \langle l \rangle = \sum_{i=1}^D p_i l_i = \sum_{i=1}^D p_i \log \frac{1}{q_i} = H(p, q),
\end{equation}
where $H(p, q)$ denotes the classical cross entropy between the distributions $p$ and $q$.

In a practical implementation where codewords must have integer lengths, we can define $l_i$ as $l_i = \lceil \log (1/q_i) \rceil$. Leveraging the properties of the ceiling function, we establish that

\begin{equation}
    H(p,q) \leq \langle l \rangle < H(p,q) + 1.
\end{equation}

This approach is known as variable-length coding, in which symbols with higher probabilities are assigned shorter codewords to optimize the overall length of the encoded message.

\subsection{A Simple Quantum Protocol}

Quantum data compression involves encoding information into qubits while considering superpositions, which introduces indeterminacy in code lengths \cite{indeter, qhuffman}. Herein, we present a novel protocol addressing this challenge. Recall that our perceived (misinformed) quantum source is $\sigma = \sum_{i=1}^D q_i |a_i\rangle \langle a_i|$, and the true source is $\rho = \sum_{i=1}^D p_i |b_i\rangle \langle b_i|$.

\textbf{Outline of the Algorithm:} We start with constructing a computational basis based on our perceived information ($\sigma$) of the quantum source. We then use a unitary transformation to map the true quantum source $(\rho)$ (which is what we actually have at hand) to the computation basis. We project the source state to the subspace with typical length, and prove that the projection is faithful. Finally, we transform the projected state back to its original basis. We show that the compression rate in this case is the quantum cross entropy $S(\rho, \sigma)$.

\textbf{Preparation Phase}: Initially, we view \( q_i \)'s as a classical probability distribution and assign a codeword \( C_i \) to the \( i \)-th symbol, with length $l_i = \log(1/q_i)$. In practice, we can only deal with integer numbers of qubits, so the precise version is $l_i = \big\lceil \log(1/q_i) \big\rceil$. This integer constraint on qubit number is noted but initially disregarded for conceptual clarity.

\textbf{Basis Construction}: As we (falsely) believe that the quantum source is $(\sigma, H)$, we construct each unit of computational basis $|i\rangle$ by assigning the first $l_i$ available qubits to $|C_i\rangle$. To keep track of a codeword's length, we define the length observable $L= \sum_{i=1}^D l_i |i\rangle \langle i|$.  When dealing with $N$ copies of the source state, the computational basis we use is $\{|i_1\rangle |i_2\rangle \cdots |i_N\rangle \}$.

\textbf{Unitary Transformation}: We define a unitary transformation \( U = \sum_{i=1}^D |i\rangle \langle a_i| \) to map the true source state \( \rho \) onto the computational basis \( |i\rangle \), yielding the transformed state \( \rho_t \) as:
\begin{equation}
    \rho_t = U \rho U^{\dagger} = \sum_{j,k} r_{jk} |j\rangle \langle k|,
\end{equation}
where $r_{jk} = \langle a_j |\rho|a_k \rangle \text{ and } r_j = r_{jj}.$

\textbf{Codeword Length}: The average length of a codeword \( \langle l \rangle \) is derived as:
\begin{equation}
\begin{aligned}
    \langle l \rangle &= \tr(\rho_t L) = \sum_i \langle a_i |\rho| a_i \rangle l_i \\
    &= \sum_i \langle a_i |\rho| a_i \rangle \log \frac{1}{q_i} = -\sum_i r_i \log q_i.
\end{aligned}
\end{equation}
By the definition of the quantum cross entropy, we have

\begin{equation}
\label{qce}
\begin{aligned}
    S(\rho, \sigma) &= -\tr (\rho \log \sigma) = - \sum_i \langle a_i | \rho | a_i \rangle \log \langle a_i | \sigma | a_i \rangle \\
    &= -\sum_i r_i \log q_i = \langle l \rangle.
\end{aligned}
\end{equation}
We observe that the quantities \( r_i \) represent the accurate probability distribution in a basis that is not optimally aligned, thus establishing a connection between quantum and classical cross entropy:
\begin{equation}
    S(\rho, \sigma) = H(r,q), \ \ r_i = \langle a_i | \rho | a_i \rangle, \ \ q_i = \langle a_i | \sigma | a_i \rangle.
\end{equation}
Here, $r$ and $q$ are probability distributions viewed in the orthonormal basis of $\sigma$.

\textbf{State for Compression}: For \( N \) identical copies of the source state, \( \rho_t^{\otimes N} \) is the state to be compressed: 
\begin{equation}
\begin{aligned}
    \rho_t^{\otimes N} 
    &= \bigg(\sum_{j_1, k_1} r_{j_1 k_1} |j_1\rangle \langle k_1| \bigg) \otimes \bigg(\sum_{j_2, k_2} r_{j_2 k_2} |j_2\rangle \langle k_2| \bigg) \otimes \\
    &\quad \cdots \otimes \bigg(\sum_{j_N, k_N} r_{j_N k_N} |j_N\rangle \langle k_N| \bigg).
\end{aligned}
\end{equation}
The cumulative length, \( l_{\text{total}} \), of the codewords is determined by summing the lengths of individual codewords. We thus define the total length observable, \( \Lambda \), as \( \Lambda = L_1 + L_2 + \cdots + L_N \). Given a basis state \( |\mathbf{x}\rangle = |i_1\rangle |i_2\rangle \cdots |i_N\rangle \), the cumulative length is 
\begin{equation}
l_{\text{total}} = \sum_{n = 1}^N \log (1/q_{i_n}).
\end{equation}
The expectation length of \( N \) codewords, \( \langle l_{\text{total}} \rangle \), is 
\begin{equation}
     \langle l_{total} \rangle = N \langle l \rangle = N S(\rho, \sigma),
\end{equation}
which is equivalent to \( N S(\rho, \sigma) \) by the definition of quantum cross entropy.

\textbf{Typicality and Projection}: 
We now demonstrate that the initial \( N S(\rho, \sigma) \) qubits encapsulate the entirety of the information within \( \rho^{\otimes N} \) in the asymptotic limit as \( N \) approaches infinity.

Let us fix $\epsilon > 0$. We define a projector $\Pi$:

\begin{equation}
    \Pi = \sum_{\text{length condition}}
    |i_1\rangle \langle i_1| \otimes
|i_2\rangle \langle i_2| \otimes \cdots \otimes
|i_N\rangle \langle i_N|,
\end{equation}
where the length condition for $\mathbf{x} = i_1, i_2, ... ,i_N$ is

\begin{equation}
    \Big|\frac{1}{N}\sum_{n=1}^{N} \log \frac{1}{q_{i_n}} - S(\rho, \sigma)  \Big| \leq \epsilon. 
\end{equation}

When $\mathbf{x}$ is $\epsilon$-strongly typical, define i.i.d random variables $I_1, I_2, ... ,I_N$ such that $I_n = \log (1/q_{i_n})$. The expectation value is $E(I) = \sum_i r_i \log (1/q_i)$. For any $\delta > 0$, when $N$ is sufficiently large, by the law of large numbers,
\begin{equation}
\begin{aligned}
    &P\left(\left|\frac{1}{N}\sum_{n=1}^{N} I_n - E(I) \right| \leq \epsilon\right) = \\
    &P\left(\left|\frac{1}{N}\sum_{n=1}^{N} \log \frac{1}{q_{i_n}} - S(\rho, \sigma)  \right| \leq \epsilon\right) \\
    &\geq 1 - \delta,
\end{aligned}
\end{equation}
which fulfills the length condition. Hence, $Q(N, \rho_t, \epsilon) \leq \Pi$, and 

\begin{equation}
    \tr(\Pi \rho_t^{\otimes N}) \geq \tr \big( Q(N, \rho_t, \epsilon) \rho^{\otimes N} \big)
    \geq 1 - \delta.
\end{equation}

We apply $\Pi$ to project $\rho_t^{\otimes N}$ onto the subspace where the total codeword length $l_{total} \in [N S(\rho, \sigma) - \epsilon, N S(\rho, \sigma) + \epsilon]$:

\begin{equation}
\gamma= \frac{   \Pi \rho_t^{\otimes N} \Pi}{\tr(\Pi \rho_t^{\otimes N})}.  
\end{equation}
We calculate the quantum fidelity to show that our data compression is indeed faithful:

\begin{equation}
\begin{aligned}
   F(\rho_t^{\otimes N}, \gamma)
   &= \left( \tr \sqrt{\sqrt{\rho_t^{\otimes N}} \gamma \sqrt{\rho_t^{\otimes N}}} \right)^2 \\
   &= \tr(\Pi \rho_t^{\otimes N}) 
   \geq 1 - \delta.
\end{aligned}
\end{equation}

When $\mathbf{x}$ doesn't satisfy the length condition, it doesn't satisfy strong typicality. When we project onto fewer qubits than $N S(\rho, \sigma)$, we miss out all the typical states, which composes the majority of all possible quantum states, and the data compression fails. Hence, $S(\rho, \sigma)$ is also the optimal compression rate under this protocol.

\textbf{Completion of Compression}: Finally, applying \( U^{\dagger} \), we revert the state to its original basis, concluding our compression protocol. 

\textbf{Connection to Quantum Machine Learning}: The compression rate of the protocol is the quantum cross entropy: $\langle l \rangle = S(\rho, \sigma)$. Our perceived quantum source state, denoted as \( \sigma \), essentially represents the model parameter. To optimize the quantum data compression protocol, we should adjust the parameter \( \sigma \) to more accurately reflect the true quantum source \( \rho \). When our information about \( \rho \) is exactly correct, we minimize the quantum cross entropy, thereby achieving the optimal compression rate, which corresponds to the von Neumann entropy.

\textbf{Consideration of Integer Qubits}: If we take into account the fact that the number of qubits has to be integer, then we have $\langle l \rangle \in [S(\rho, \sigma), S(\rho, \sigma) + 1)$, and we need no more than $N S(\rho, \sigma) + N$ qubits for a successful data compression even when our knowledge of the quantum source is wrong. Of course, if our perceived source state $\sigma$ is far from the true source state $\rho$, $S(\rho, \sigma)$ is huge and we will be better off by just sending all the information-bearing qubits, which gives a compression rate of $\lceil \log D \rceil$ qubits. 

The novelty of the algorithm lies in the fact that we incorporate ideas from classical transform compression by generalizing the concept of transforming a source's data into a format that can be more efficiently compressed. We also introduce typical length, which is an improvement from Schumacher compression \cite{qcoding}, where the typicality projection is too strict. These methods make our protocol more adaptive and can be applied to a universal class of sources, namely anything that could be represented by a quantum density matrix. The protocol ensures that the essential information is retained by leveraging quantum mechanical principles, like superposition and entanglement, and quantifying information content through quantum cross entropy.

\ 

\section{Conclusion}

In this work, we draw connections between quantum machine learning and quantum data compression. In particular, we present an operational interpretation of the quantum cross entropy, that the quantum cross entropy is the compression rate for sub-optimal quantum source coding. To achieve this, we introduce a simple, novel, and universal quantum data compression protocol. Evaluating the time and query complexity \cite{complexity2, complexityunitary, shangnan2019complexity, querycomplexity} required to implement this protocol on quantum computers is a fascinating prospect.
With a quantitative measure like complexity, it enables us to compare this protocol with other quantum compression protocols and find improvements. Ultimately, a broader and deeper understanding of the quantum cross entropy can guide us in designing efficient quantum machine learning algorithms, which leads to solutions to more challenging problems in quantum information.

\begin{acknowledgements}
We thank Patrick Hayden, Stephen Shenker, and Leonard Susskind for insightful discussions on the work. We also thank helpful comments from Anurag Anshu, Adam Bouland, Xie Chen, Xun Gao, Tarun Grover, Russell Impagliazzo, John McGreevy, David Meyer, Jim Kurose, Shachar Lovett, Mikhail Lukin, Joel E. Moore, Brian Swingle, Don Towsley, and Umesh Vazirani.
This work is supported by the Simons Foundation.
\end{acknowledgements}

\bibliography{apssamp}
\end{document}